\documentclass{elsart3p}

\usepackage{amssymb}
\usepackage{epsfig}

\def\av{{\sc Antares/Virgo}}
\def\anta{{\sc Antares}}
\def\vo{{\sc Virgo}}
\def\lo{{\sc Ligo}}
\begin{document}

\begin{frontmatter}



\title{Coincidences between Gravitational Wave Interferometers \\ and High Energy Neutrino Telescopes}


\author{Thierry PRADIER}

\address{{\sl Institut Pluridisciplinaire Hubert Curien} {\sc (iphc/drs)} \\
\&
\\University Louis-Pasteur, Strasbourg (France)}

\begin{abstract}
Sources of gravitational waves (GW) and emitters of high energy (HE) neutrinos both involve compact objects and matter moving at relativistic speeds. GW emission requires a departure from spherical symmetry, which is the case if clumps of matter are accreted around black holes or neutron stars, and ejected in relativistic jets, where neutrinos are believed to be produced. 
Both messengers interact weakly with the surrounding matter, hence point directly to the heart of the engines that power these emissions. Coincidences between GW interferometers ({\it e.g.} \vo) and HE $\nu$ telescopes ({\it e.g.} \anta) would then give a unique insight on the physics of the most powerful objects in the Universe. 
The possibility, observability and detectability for such GW/HE $\nu$ coincidences are analysed.
\end{abstract}

\end{frontmatter}

\section{Scientific motivations}
\label{sec:intro}

The forthcoming years should be very exciting both in gravitational wave (GW) astronomy and high energy neutrino 
(HE $\nu$) astronomy. The \vo~interferometer~\cite{virgo}, currently closed down for upgrade, should be taking 
data with an improved injection system in 2009, whereas the \anta~collaboration has completed the deployment 
and connections of its 12 lines~\cite{antares} early in June 2008, starting the operation of the first {\it undersea} 
neutrino telescope. In the mean time, the two \lo~interferometers (ITF) are in operation~\cite{ligo}, and {\sc IceCube} 
is deploying its 1~km$^3$ neutrino telescope in the ice of the South Pole, already having 40 lines in data taking 
mode~\cite{icecube}. 

Both GW sources and HE $\nu$ emitters involve compact objects and matter moving at relativistic speeds. As a result, coincidences between GW interferometers and neutrino telescopes can be envisaged~\cite{coinc}, and should then be feasible 
by 2009, with the joint operation of \vo, \lo, \anta~and {\sc IceCube}. Together, they would give a unique insight on the
 physics of the most powerful objects in the Universe. Some classes of astrophysical object, invisible in electromagnetic
 channels, may be observable only {\it via} their  gravitational waves and high energy neutrino emissions. 
Finally, in many quantum gravity (QG) models~\cite{qg}, the propagation velocity of a particle depends on the 
energy:

\begin{equation}
c^2 p^2 = E^2 \left[1 + \xi \left( \frac{E}{E_{\textrm{{\tiny QG}}}} \right) + {\mathcal O}\left( \frac{E^2}{E_{\textrm{{\tiny QG}}}^2} \right)+ \ldots \right]
\label{eq:intro_qg1}
\end{equation}

\noindent
where $\vert \xi \vert \simeq 1$, and $E_{\textrm{{\tiny QG}}}$ is the energy scale at which QG effects arise. Hence, measuring a non-zero time delay between gravitational wave bursts and high energy neutrino transients would then allow to probe Quantum Gravity effects at the Planck energy level.

Such time coincidences require GW bursts (localized in time), and would allow GW antennae to confirm a burst detection, and neutrino telescopes to sign the cosmic origin of the signal.
It must be first demonstrated that such a coincidence process is:

\begin{itemize}
\item[$\bullet$] possible: sources of both GW and HE $\nu$, able to emit signals possibly coincident in time, exist in the Galaxy. This is discussed in section~\ref{sec:sources};
\item[$\bullet$] observable: the visibility sky maps of \vo~and \anta~are not orthogonal. This point is addressed in section~\ref{sec:obs} ;
\item[$\bullet$] detectable: the resulting coincidence detection probability, for a fixed accidental coincidence rate, which constrains the authorized background level for each individual detector, must be estimated. This point is developed in section~\ref{sec:detec}.
\end{itemize}

\section{GW bursters and HE $\nu$ sources}
\label{sec:sources}

Only {\it galactic} potential sources of both GW and HE $\nu$ signals will be discussed here, for two reasons.
 Firstly, the only sources accessible with first generations detectors such as \anta~and \vo~are likely to 
be galactic ones. Secondly, for sources in the local universe, in particular with negligible redshift, the 
QG delay mentionned in the previous section is independent of cosmological models. This (non-exhaustive) section 
focuses on microquasars and soft-gamma repeaters, however core-collapse supernovae and gamma-ray bursts are commonly cited extra-galactic sources of both GW and $\nu$ emissions: these are described elsewhere~\cite{sngrb}.

\subsection{Outbursts from microquasars}

Microquasars are galactic jet sources associated with
some classes of X-ray binaries involving both neutron
stars and black hole candidates. During
active states, the X-ray flux and spectrum can vary
substantially, with a total
luminosity that, during the so-called very high states,
often exceeds the Eddington limit~\cite{mirabel}. Their activity involves
ejection within jets with kinetic power that appears to constitute
a considerable fraction of the liberated accretion
energy, giving
rise to intense radio and IR flares. Radio monitoring of
some X-ray transients has revealed superluminal motions
in some objects, indicating that the jets are relativistic, with $\gamma\sim 1-10$.
The ejection episodes
are classified into several classes according to the brightness
of synchrotron emission produced in the jet and the
characteristic time scale of the event. The duration
of major ejection events is typically on the
order of days, while that of less powerful flares is correspondingly shorter (minutes to hours).
The correlations between the X-ray and synchrotron
emission clearly indicates a connection between the accretion process and the jet activity. Whether radio and
IR outbursts represent actual ejection of blobs of plasma
or, alternatively, formation of internal shocks in a quasi-steady
jet is unclear. In any case, since
the overall time scale of outbursts is
much longer than the dynamical time of the compact
object (milliseconds), it is likely that shocks will continuously
form during the ejection event. If a fraction
of at least a few percent of the jet power is used to accelerate electrons to very high energies then
emission of high-energy gamma rays is anticipated, in
addition to the observed radio and IR emission.

\subsubsection{High energy neutrino emission}

The content of jets in microquasars remains an open issue.
In scenarios in which an initial rise
of the X-ray flux leads an ejection of the inner part of the
accretion disk, as is widely claimed to be suggested by the
anticorrelation between the X-ray and radio flares seen during
major ejection events,
e-p jets are expected to be produced. A
possible diagnostic of e-p jets is the presence of Doppler-shifted
spectral lines, such as the H${\alpha}$ line as seen in SS433.
\begin{figure}[h!]
\centerline{\includegraphics[width=\linewidth]{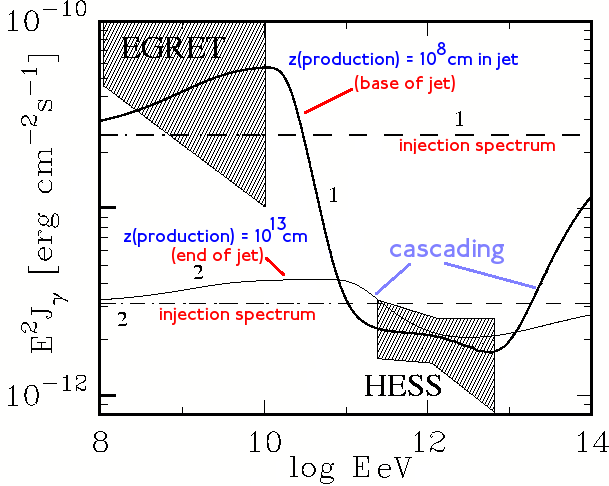}}
\caption{$\gamma$ Data from EGRET and HESS, together with predictions for models with production of $\gamma$ at the base or at the end of the jet. 
\label{fig:microq_henu}}
\end{figure}

Taking the example of LS 5039, some authors argue in favor of hadronic origin of TeV photons~\cite{ls5039}, especially if they are produced within
the binary system. The detected 
$\gamma$-rays should be accompanied by a flux of high energy
neutrinos emerging from the decays of $\pi^{\pm}$ mesons produced at pp and/or p$\gamma$
 interactions. The
flux of TeV neutrinos, which can be estimated on the basis of the detected TeV 
$\gamma$-ray flux, taking
into account the internal $\gamma\gamma\rightarrow e^+e^-$ absorption, depends significantly on the location of $\gamma$-ray
production region. The minimum neutrino flux above 1 TeV is expected to be at the level
of 10$^{−12}$~cm$^{−2}$s$^{-1}$, but could be up to a factor of 100 higher. As can be seen in figure~\ref{fig:microq_henu} (taken from~\cite{ls5039}), the HESS/EGRET data agree well with a production of $\gamma$ (and neutrinos) at the base of the jet, very close to the onset of the acceleration phase and its corresponding GW signal (see section~\ref{sec:gwmicroq}).
Finally, one should note that the detectability by \anta~or future km$^3$ telescopes depends strongly on the high energy cutoff in the spectrum
of parent protons.

\subsubsection{GW emission during a powerful flaring event}
\label{sec:gwmicroq}

Two kinds of processes could lead to detectable GW signals~\cite{mypaper}. Firstly, the matter accreted for months/years could be {\it swallowed} by the compact object, and, provided that the process is fast, trigger the resonance of normal modes in the central object. Secondly, the acceleration of the matter in the jet is the origin of a short GW burst.

For both signals, the amplitude will depend critically on the accreted/ejected mass. The assumption which is made in the following is that all the matter ejected during a flare comes from the accretion disk, and had previously fallen onto the compact object at some previous moment.  The GW emission produced by the infall of matter onto the central object is different in the case of neutron stars (NS) and black holes (BH), but in both cases results in the excitation of the Quasi-Normal Modes of the star~\cite{gwbh} (typically a damped sine signal) which could continue into the ejection phase. The time-lag between the two processes is unknown, and could range from ms up to several days.

\begin{figure}[h!]
\centerline{\includegraphics[width=\linewidth]{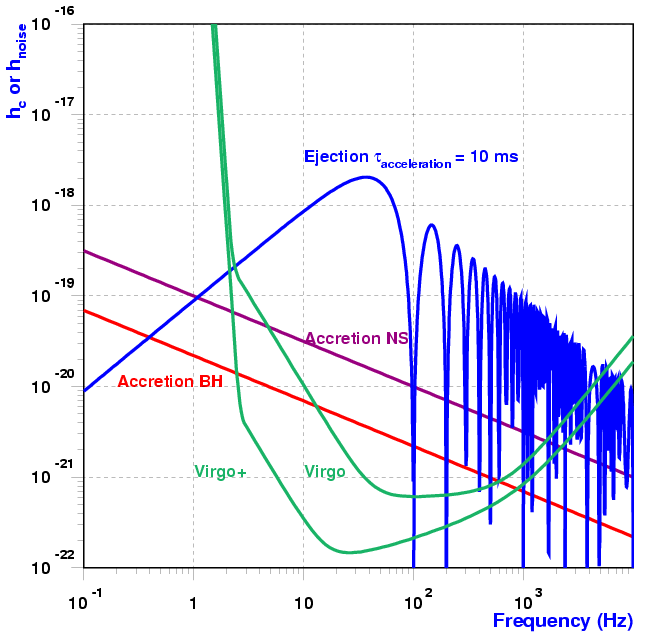}}
\caption{Amplitudes expected for the accretion/ejection processes at 1~kpc, for $\delta m \sim 10^{-4} \textrm{M}_{\odot}$, $\tau_{\textrm{acc}} \sim 10$~ms, $\gamma = 10$.
\label{fig:microq_gw2}}
\end{figure}
In the NS case, the characteristic amplitude (scaling as $1/d$) at 1 kpc  can be written as $h_c\approx 10^{-20} \left(\frac{\eta}{10^{-6}}\right)^{1/2} f_0^{-\frac{1}{2}}$ for the accretion signal~\cite{gwbh}, to be compared with the \vo~sensitivity $h_{\textrm{{\tiny noise}}} \sim 10^{-21}$ around 1 kHz. $\delta m = \eta \textrm{M}_{\odot}$ is the accreted/ejected mass, and $f_0$ the frequency of the excitation. Note that this assumes a sudden infall of the matter onto the compact object ({\it bursting mode}), and does not stand if the accretion infall is a continuous and slow process ({\it continuous mode}).

The acceleration of an ultrarelativistic blob of matter with a Lorentz factor $\gamma$ around a compact object induces a {\it burst with memory}, the space-time perturbation reaching a maximum amplitude at 1 kpc $\Delta h\sim 2\times10^{-22} \left(\frac{\gamma}{10}\right) \left(\frac{\eta}{10^{-6}}\right)$, independent on the density of the ejecta~\cite{jetgw}. The frequency is typically the inverse of the acceleration time $t_{\textrm{{\tiny acc}}}$, ranging from $\Delta t_{\textrm{{\tiny free-fall}}} \sim 0.1$~ms for a 10 M$_{\odot}$ object, up to $\Delta \tau_{\textrm{{\tiny max}}} \sim 1$~min, observations of galactic X-Ray binaries showing that radio emission occurs at a distance $\sim 0.1$ AU from the central object. A summary of those estimates are displayed in figure~\ref{fig:microq_gw2}, for a source at 1~kpc, ejected mass $\delta m \sim 10^{-4}~\textrm{M}_{\odot}$ and acceleration time $\tau_{\textrm{{\tiny acc}}} \sim 10$~ms.

Taking into account accretion rates (typically $\dot{M}\sim 10^{-8} M_{\odot}/$yr), jet luminosities ($10^{29} \textrm{J/s} - 10^{33} \textrm{J/s}$) and radio blob sizes $\sim10^{12}$~m and densities (from $10^{-10} \textrm{g.cm}^{-3}$ up to $10^{15} \textrm{g.cm}^{-3}$), the accreted/ejected mass for major events can be estimated to range from $10^{-8} \textrm{M}_{\odot}$ up to $10^{-4} \textrm{M}_{\odot}$, with an average mass of $10^{-6} \textrm{M}_{\odot}$. Thus the amplitudes shown in figure~\ref{fig:microq_gw2} correspond to the maximum mass and Lorentz factor that can be expected. Nonetheless, such extreme outbursts, where the matter accreted over months or even years, is swallowed and then ejected in one single bursting event cannot be ruled out.

\subsection{Flares from Soft-Gamma Repeaters}

Soft gamma-ray repeaters (SGRs) are X-ray pulsars which have quiescent soft (2-10 keV) periodic X-ray emissions with periods ranging from 5 to 10 s. They exhibit repetitive bursts lasting $\sim$ 0.1 s which reach peak luminosities of $\sim 10^{34}$ J/s, in X-rays and $\gamma$-rays. There are 4 known SGRs, 3 in the Milky Way, and one in the Large Magellanic Cloud. Three of the 4 have had hard spectrum (MeV energy) giant flares with luminosities up to $10^{40}$ J/s. The favoured {\it magnetar} model for these objects is a neutron star with a huge magnetic flied $B\sim10^{15}$ G~\cite{sgr}. Star-quakes are thought to fracture the rigid crust causing outbursts. These giant flares result from the formation and dissipation of strong localized currents coming from magnetic field rearrangements associated with the quakes, liberating a high flux of X- and $\gamma$-rays.

Sudden changes in the large magnetic fields would accelerate protons or nuclei that produce neutral and charged pions in interactions with thermal radiation. These subsequently decay into  TeV or even PeV energies $\gamma$-rays and neutrinos~\cite{sgrnu}. Flares from SGRs are thus potential sources of high energy neutrinos.

During the crustal disruption, a fraction of the initial magnetic energy is annihilated and released as photons, and the stored elastic energy is also converted into shear vibrations with frequencies in the kHz regime. These waves are able to excite non-radial modes, damped by GW emission~\cite{sgrgw}. The expected gravitational strain amplitude can be written as:

\begin{equation}
h(t) = \frac{2}{d \omega_n} \left(\frac{G E}{c^3 \tau_n}\right)^{\frac{1}{2}} e^{\left(i\omega_n t - \frac{t}{\tau_n} \right) },
\label{eq:h_sgr}
\end{equation}

\noindent 
where $E$ is the total energy, $G$ the gravitational constant, $\omega_n$ and $\tau_n$ the pulsation and damping timesale of the $n$-mode. These oscillation parameters depends on the equation of state and the stellar mass. Simulations~\cite{sgrgw} show that low mass
stars produce larger gravitational amplitudes and could be detected
more deeply within the Galaxy. Moreover, sources at distances ranging from 0.4 kpc up to 2.4 kpc could be probed with the planned sensitivity of \vo. The detection probability and frequency of these events depend on the poorly constrained distribution of this class of sources in the Galaxy.

\section{Observability of Coincidences}
\label{sec:obs}

Both GW detectors and HE $\nu$ telescopes have limited sky coverage and exposure. In order to perform coincidences between both types of detector, the overlap of such visibility maps has to be computed: the following paragraphs address this question, taking the examples of \vo~and \anta.

\subsection{\vo~beam pattern}

\begin{figure}[h!]
\centerline{\includegraphics[width=\linewidth]{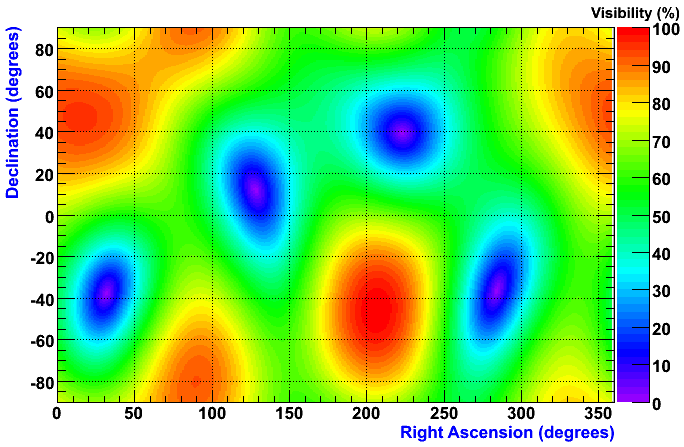}}
\caption{\vo~beam pattern in equatorial coordinates.
\label{fig:visu_virgo}}
\end{figure}

The response $h(t)$ of an interferometric detector to a GW is a linear combination of the two independent wave polarizations $h_+$ and $h_{\times}$, with weighting factors called the beam pattern functions: they have values in the range $[-1;1]$, depending on the longitude and latitude of the detector location, as well as its orientation, the angle between the arms, the sky coordinates of the source, and the wave polarization angle. The best response is achieved for detectors with orthogonal arms. In the following, average over the unknown polarization angle will be presented. The instantaneous beam pattern (normalised to its maximal value) at a given time during the day, in equatorial coordinates (right ascension $\alpha$, declination $\delta$), is displayed in figure~\ref{fig:visu_virgo} for \vo. 

\subsection{\anta~visible sky}

\begin{figure}[h!]
\centerline{\includegraphics[width=\linewidth]{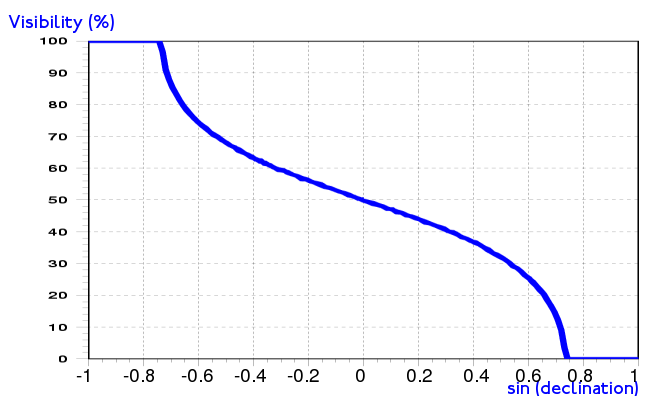}}
\caption{\anta~daily averaged visibility {\it vs} $\sin \delta$.
\label{fig:visu_antares}}
\end{figure}

\anta~is only sensitive to sources below the horizon at some time during the day, because it searches for neutrinos 
which have interacted in the Earth: a portion of the sky is therefore never visible. Figure~\ref{fig:visu_antares} 
shows the daily average visibility as a function of $\sin \delta$.

\subsection{\av~common sky}

\begin{figure}[h!]
\centerline{\includegraphics[width=\linewidth]{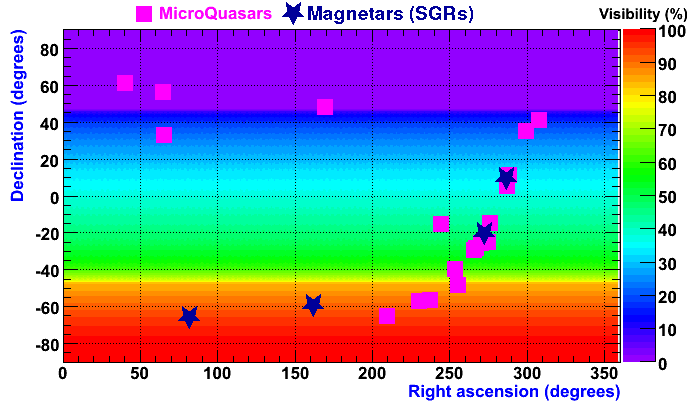}}
\caption{Common visibility sky map for \anta~and \vo.
\label{fig:visu_common}}
\end{figure}

The visibility sky map for coincidences between \anta~and \vo~is the convolution of the two previous exposure maps. The daily averaged common sky map is displayed in figure~\ref{fig:visu_common}, together with the position of known microquasars and soft-gamma repeaters (or magnetars). Except for three of them, all are visible at some time by both experiments, rendering observable GW/HE $\nu$ coincidences for most of these galactic sources.

\section{Detectability of Coincidences}
\label{sec:detec}

Given GW/HE $\nu$ coincidences are observable, at least for the \anta~and \vo~couple, the detectability, in terms of detection probability {\it vs} false alarms rate, has to be evaluated.

\subsection{Coincidence window and QG effects}

To set the coincidence time window, possible physical propagation delays have to be estimated. In the case of GW, 
the graviton being massless, and the energy carried away by each individual graviton in a GW burst being small 
($E_{\textrm{{\tiny graviton}}}\sim hf\ll 1$ for $f = 1~\textrm{kHz}$), QG-induced or mass-induced delays are 
close to zero. For a 1 TeV $\nu$, the mass-induced delay is negligible even with $m_{\nu} = 1~\textrm{eV}$. 
Taking the expression given in eq.~\ref{eq:intro_qg1} as a starting point, with $\xi=-1$ being favoured, 
and neglecting any cosmological effects (for low redshift $z \ll 1$), the delay in ms becomes, in the first order:

\begin{equation}
\Delta t^{\textrm{{\tiny ms}}}_{\textrm{{\tiny QG}}} \simeq 0.15 \left(\frac{d}{10~\textrm{kpc}} \right) \left(\frac{E_{\nu}}{1~\textrm{TeV}}\right)\left(\frac{10^{19}~\textrm{GeV}}{E_{\textrm{{\tiny QG}}}}\right)
\label{eq:intro_qg2}
\end{equation}

Taking $E_{\textrm{{\tiny QG}}} = E_{\textrm{{\tiny Planck}}} \sim 10^{19}~\textrm{GeV}$, this yields a maximum 
QG delay of 1 second for $E_{\nu}=1~\textrm{PeV}$ and sources up to the Large Magellanic Cloud ($d \sim 50~\textrm{kpc}$),
 or lower energy neutrinos ($E_{\nu} \approx 1~\textrm{TeV}$) and sources as far as the Virgo Cluster 
($d \sim 20~\textrm{Mpc}$). $\Delta t_{\textrm{{\tiny coinc}}} = 1~$s thus seems a reasonnable choice. 
Nevertheless, the coincidence time window can also be set by imposing an overall coincidence detection 
probability for a given GW signal. Detection issues, both for GW ITFs and HE $\nu$ telescopes must now be addressed.

\subsection{\vo~detection}

\begin{figure}[h!]
\centerline{\includegraphics[width=\linewidth]{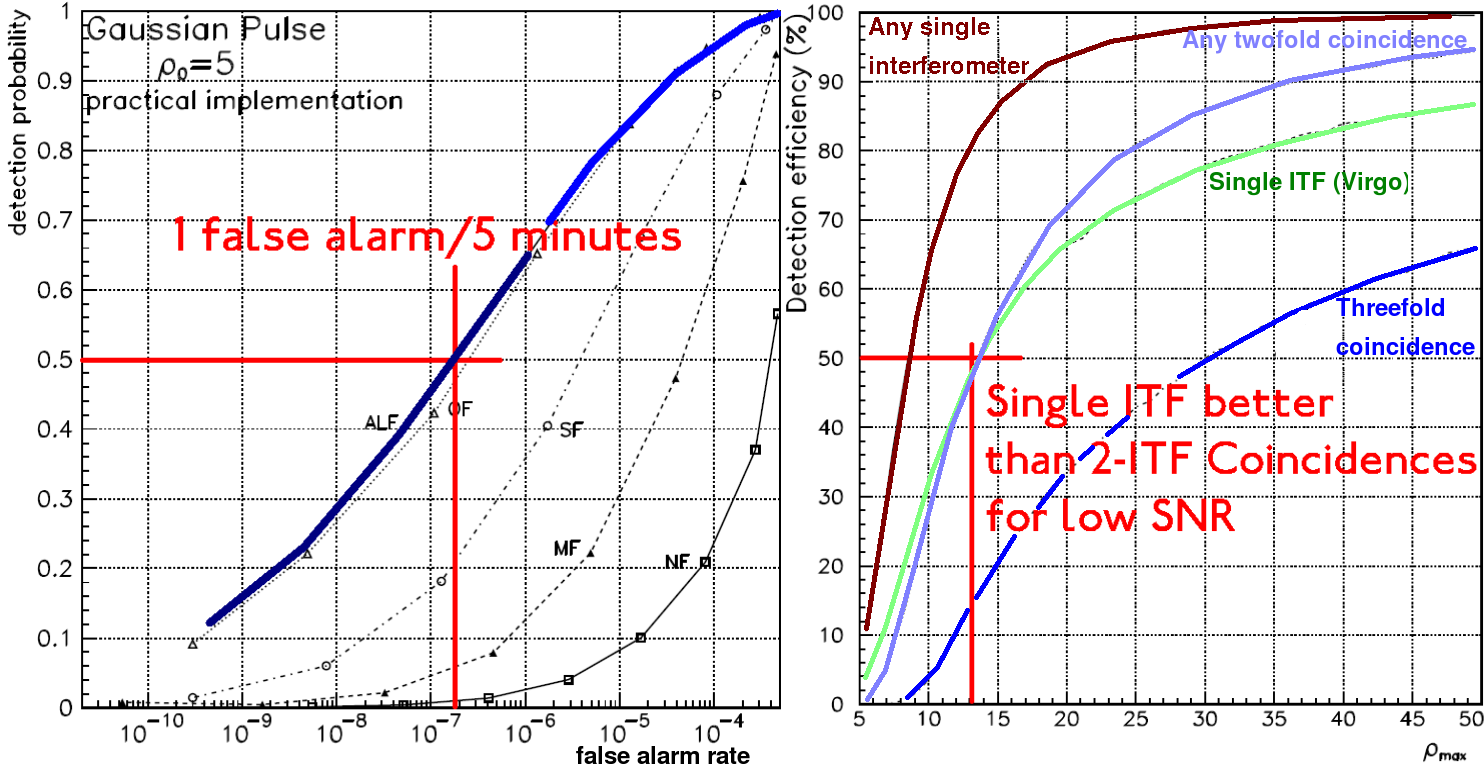}}
\caption{Left : detection probability {\it vs} false alarm rate, for a $\rho = 5$ gaussian pulse. Right : detection probability {\it vs} $\rho_{\textrm{{\tiny max}}}$ for the different possible detection configurations.
\label{fig:detection_virgo}}
\end{figure}

The detection of a transient signal in a GW ITF is not an easy task. Generally, an allowed false alarm rate is fixed, 
and the detection probability can be estimated as a function of the signal-to-noise ratio (SNR or $\rho$) of a particular 
signal: this is shown for \vo~in figure~\ref{fig:detection_virgo}, for $\rho_{\textrm{{\tiny max}}} = 5$ (defined as 
the SNR obtained with optimal detector orientation and perfect detection by Wiener filtering), a very low signal, 
for detection algorithms designed for burst signals~\cite{virgodetection2}. In the best case, a threshold corresponding 
to 1 false alarm every 5 minutes is needed to obtain a 50\% detection probability, without taking into account any 
beam pattern effects.

The right part of figure~\ref{fig:detection_virgo} displays the detection efficiency as a function of $\rho_{\textrm{{\tiny max}}}$, 
for this particular false alarm rate, in the case of a single ITF detection or coincident detection in the 
\vo/\lo~network: for low SNRs, the detection by a single ITF is more probable than any twofold coincidence, 
and the detection by any single ITF is always more efficient than any coincidence configuration~\cite{virgodetection}. 
In the case of the detection by any of the 3 ITFs, the directional information is not available (no triangulation), 
and the only relevant information is therefore the {\it time} of the burst event.

\vo~has a sampling frequency of 20 kHZ, and for a gaussian burst 
of width $\tau$ and SNR $\rho$, the rms error on the burst arrival time is~\cite{virgodetection}:

\begin{equation}
\Delta t^{\textrm{{\tiny RMS}}} \approx \frac{1.5}{\textrm{SNR}} \left(\frac{\tau}{1~\textrm{ms}} \right)~\textrm{ms},
\label{eq:gw_time}
\end{equation}

\noindent
yielding a timing resolution below the ms for $\rho > 5$ and short burst. This of course limits the accessible 
QG energy scale, and the coincidence window to be used.

\subsection{\anta~detection}

\begin{figure}[h!]
\centerline{\includegraphics[width=\linewidth]{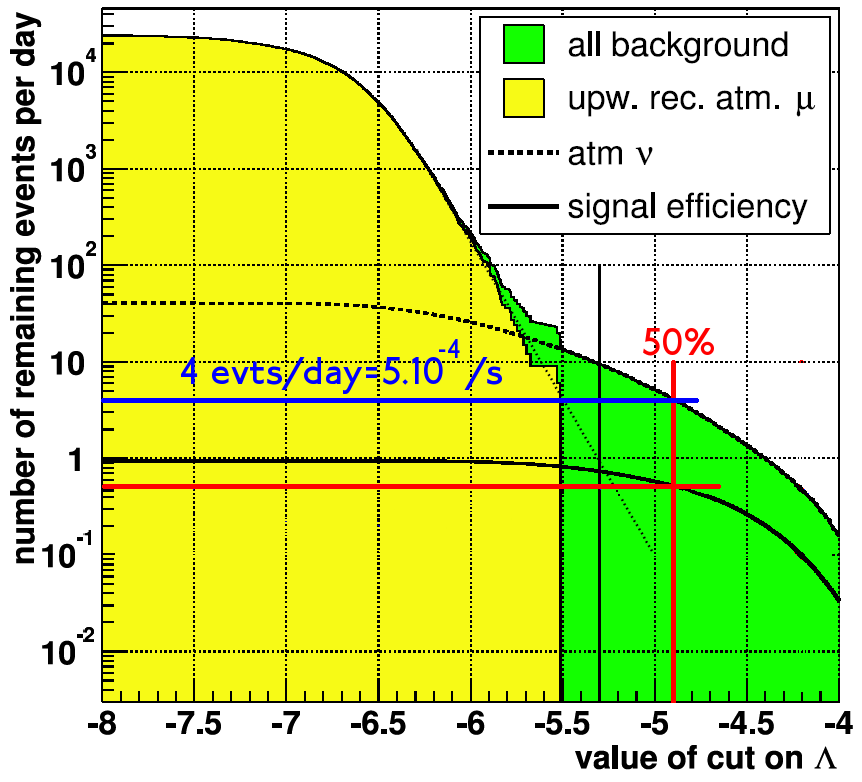}}
\caption{Number of background events left as a function of the cut value for $\Lambda$. The efficiency for signal is also shown.
\label{fig:detection_antares}}
\end{figure}

In a neutrino Telescope, the \v{C}erenkov light emitted
by the neutrino-induced muon is detected by an array of photomultipliers arranged in strings, able to reconstruct the energy and 
direction of
the incident muon/neutrino. The measurements of the time of the hits and the amplitude of the hits, together 
with the position of the hits are needed to achieve the reconstruction of the muon track with the desired 
resolution (below 0.3$^{\circ}$ above 10 TeV). The quality of the track fit is often expressed in terms of a 
log-likelihood ratio term $\Lambda \approx \frac{\log(\mathcal{L})}{N_{\textrm{dof}}}$, the distribution of which 
is shown in figure~\ref{fig:detection_antares}, for (upward) atmospheric neutrinos and misreconstructed atmospheric muons, together with the signal detection efficiency ({\it i.e.} the ability to 
detect within 1$^{\circ}$ of the true direction a signal neutrino, assuming a $E^{-2}$ spectrum). The standard cut 
applied is $\Lambda = -5.3$, for which the signal efficiency is close to 75\%; the misreconstruted atmospheric 
muons are decreased to 1/day, and the atmospheric neutrinos to 10/day~\cite{aart}. 

\subsection{Accidentals and efficiency}

\begin{figure}[h!]
\centerline{\includegraphics[width=\linewidth]{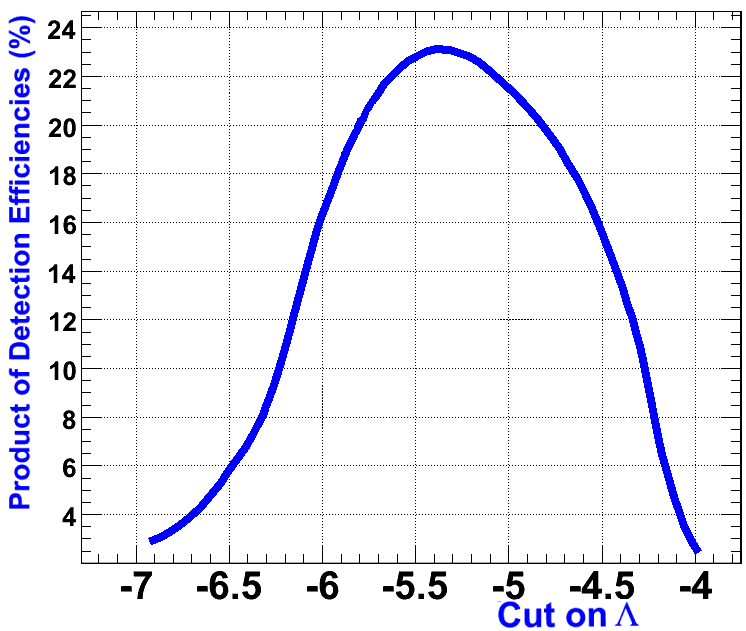}}
\caption{Coincidence detection probability {\it vs} $\Lambda$ cut.
\label{fig:result1}}
\end{figure}

Figures~\ref{fig:detection_virgo} and~\ref{fig:detection_antares} provide the information needed to estimate 
the detection probability $\epsilon_{{\tiny V,A}}$ for a background/false alarm level $R_{{\tiny V,A}}$ in both 
detectors (V for \vo, A for \anta). The coincidence detection probability is 
$\epsilon_{\textrm{{\tiny coinc}}} = \epsilon_{{\tiny V}} \epsilon_{{\tiny A}}$, whereas the coincident accidentals 
rate in a given time window is $R_{\textrm{{\tiny coinc}}} = R_{{\tiny V}} R_{{\tiny A}} \Delta t_{\textrm{{\tiny coinc}}}$. 
Setting $\Delta t_{\textrm{{\tiny coinc}}} = 1$~s and $R_{\textrm{{\tiny coinc}}} \sim 1/$yr, 
the resulting coincidence detection probability is shown in figure~\ref{fig:result1} as a function of the 
$\Lambda$ cut: the efficiency is maximum for $\Lambda \sim -5.5$, below which $R_{{\tiny A}}$ is too high, 
resulting in a too high \vo~detection threshold (for a preset $R_{{\tiny V}}R_{{\tiny A}}$).

\begin{figure}[h!]
\centerline{\includegraphics[width=\linewidth]{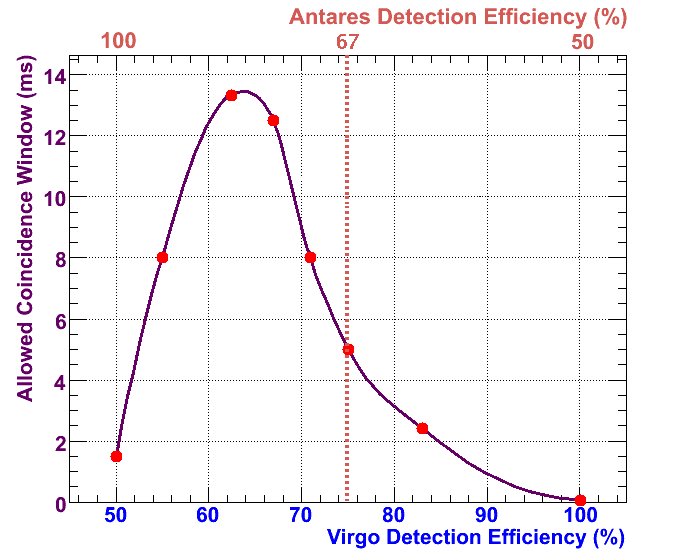}}
\caption{Maximal allowed coincidence time window in ms vs \anta/\vo~detection probabilities, requiring $\epsilon_{\textrm{{\tiny coinc}}} =50 \%$.
\label{fig:result2}}
\end{figure}

Equivalently, the coincidence detection probability can be set at {\it e.g.} 50\% for a given signal, and 
the maximal allowed coincidence time window can be extracted: this is displayed in figure~\ref{fig:result2}, 
as a function of \anta/\vo~detection probabilities, for a $\rho=5$ gaussian burst, with 
$R_{\textrm{{\tiny coinc}}} \sim 1/$yr. The time coincidence window is maximal for $\epsilon_{{\tiny V}} \sim 65\%$, 
reaching $\sim 15$~ms. 

For such a low accidental rate (1/yr), several coincident detections would be needed to have a high significance, 
but this nonetheless proves the detectability of GW/HE $\nu$ coincidences.

\section{\av~Coincidences}
\label{sec:coinc}

If a coincidence is indeed observed, the significance of this positive detection can be estimated, and assuming that 
both the GW and HE $\nu$ signals have been emitted with zero delay at the source, limits can be put on the QG energy 
scale $E_{\textrm{{\tiny QG}}}$.
In the case described in the previous section (requiring $\epsilon_{\textrm{{\tiny coinc}}} = 50 \%$), the minimum 
accessible $E_{\textrm{{\tiny QG}}}$ can be estimated using eq.~\ref{eq:intro_qg1}. The maximum energy scale yielding 
a measurable effect is limited by GW timing resolution, which depends on the burst duration and SNR, and reaches in 
this case $E^{\textrm{{\tiny max}}}_{\textrm{{\tiny QG}}} \sim 5\times10^{18}$~GeV ($\rho=5$), close to the Planck limit. 
The minimum accessible energy scale is in turn determined by the maximal coincidence window defined previously, which 
yields $E^{\textrm{{\tiny min}}}_{\textrm{{\tiny QG}}} \sim 10^{17}$~GeV. This is to be compared with existing limits 
on $E_{\textrm{{\tiny QG}}}$, {\it e.g.} using TeV flares from Mrk421 $\sim 4\times10^{17}$~GeV~\cite{qgexp}. It should 
be noted that to perform a real {\it measurement} of $E_{\textrm{{\tiny QG}}}$, the neutrino energy resolution is of 
importance, and is a factor 2 or 3 in the case of \anta~\cite{aart}.

\subsection{Making coincidences}

The process for performing time coincidences is quite classical. \vo~and \anta~produce 
trigger lists, according to some predefined false alarm rates, corresponding to different 
coincidence time windows. Timeshifts performed on these data streams allow for the study of
 background coincident triggers; in the {\it zero-lag} case, \anta/\vo~triggers are compared 
in predefined time windows. The significance of an observed coincidence is then statistically 
evaluated by comparing the two cases.

\subsection{\av~ common calendar}

\begin{figure}[h!]
\centerline{\includegraphics[width=\linewidth]{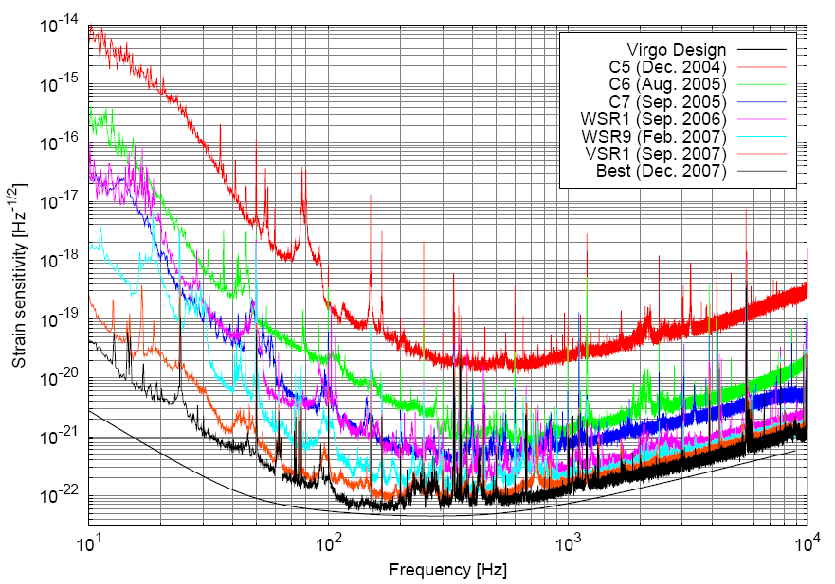}}
\caption{\vo~sensitivity curves from Commissioning Runs to Science Runs.
\label{fig:virgosens}}
\end{figure}

\vo~took data jointly with the 2 \lo~interferometers between May and September 2007, 
during the {\it Virgo Scientific Run} (VSR), achieving the sensitivity shown in figure~\ref{fig:virgosens}, 
together with the sensitivities obtained during previous commissioning or science runs (weekly runs from 
september 2006 until march 2007). In spite of problems with the laser injection system (at low frequency), 
and a factor 2 difference at high frequency, the resulting sensitivity is quite close to the expectation. 
The interferometer should be taking data again with an improved injection system in 2009.

\anta~has been taking data continuously with its final 12 line configuration since the end of May 2008, and is expected to observe high energy neutrinos for a period of 10 years. Interestingly, during the {\it VSR}, \anta~already had 5 lines operational since January 2007. Clearly, during this parallel operation of \vo~and \anta~5-lines, only the most powerful GW/HE $\nu$ sources could be detected; however these data could be used as a {\it test bench} for preliminary studies on time coincidences. Moreover, \lo~data, 
which are also available for this {\it VSR}/5-lines period, could be used to enhance the detection probability, 
a detection by any of the 3 interferometers being more efficient than for \vo~alone (see section~\ref{sec:detec}).

\section{Conclusions}

Time coincidences between GW interferometers, such as \vo, currently down for upgrade, and high energy neutrino 
telescopes, such as \anta, now fully operational, are thus feasible: common sources able to emit coincident signals 
in both channels exist in our own galaxy. Microquasars, during major outbursts, and flares from SGRs are possible targets. 
\anta~and \vo~visibility sky maps are not othogonal, allowing for coincident detections. Taking into account 
the relationship between detection efficiencies and false alarm rates in \anta~and \vo, coincident observations appear possible. 

Such coincidences can be performed using \anta~5-lines and {\it VSR} data (end of 2007) as a test that could be performed before the 
upgrade of \vo~next year, which should improve the sensitivity at low and high frequency to reach the design sensitivity. 
This {\sc Virgo+} upgrade will correspond with the {\it routine} operation of the full \anta~detector, provinding the 
opportunity to perform the time coincidences presented in this paper.

\begin{figure}[h!]
\centerline{\includegraphics[width=\linewidth]{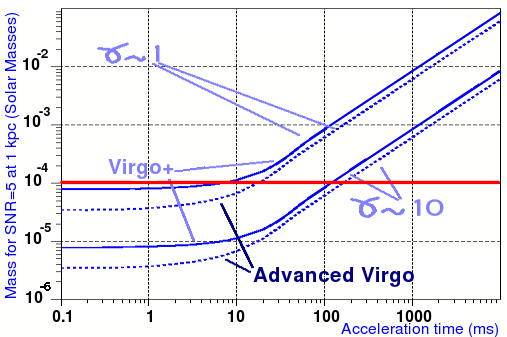}}
\caption{Ejected mass needed for SNR=5 for a microquasar at 1 kpc for {\sc Virgo+} and {\sc Advanced Virgo}.
\label{fig:adv_virgo}}
\end{figure}

Finally, {\it circa} 2015, a km$^3$ neutrino telescope should be operating in the Mediterranean Sea~\cite{km3}, along with 
an {\sc Advanced Virgo} interferometer~\cite{adv_virgo}, with enhanced sensitity at low frequency. Figure~\ref{fig:adv_virgo}~\cite{mypaper} 
shows the ejected mass needed in a microquasar ejection event to obtain a SNR = 5 in {\sc Virgo+} and 
{\sc Advanced Virgo}~\cite{adv_virgo} as a function of the acceleration time (see section~\ref{sec:sources}): 
less extremes ejection scenarios could be probed, and interesting constraints on accretion/ejection models could 
be set by this novel multimessenger approach.

\noindent
\textbf{Aknowledgements : } \textit{I would like to thank the Organizing Committee, especially John Carr, for trusting 
me for this plenary talk.}


\begin{thebibliography}{21}

\bibitem{virgo} \vo~Collaboration, Nucl. Phys. Proc. Suppl. \textbf{54B} 167-175 (1997); see also : \vo, {\it AIPC} \textbf{924}, 187-193 (2007)
\bibitem{antares} M. Circella, for \anta, these proceedings
\bibitem{ligo} \lo~Collab., Class. Quantum Grav. \textbf{25} 114041~(2008)
\bibitem{icecube} E. Resconi, for {\sc IceCube}, these proceedings
\bibitem{coinc} Y. Aso {\it et al.}, Class. Quantum Grav. \textbf{25} 114039~(2008)
\bibitem{qg} S. Choubey \& S. F. King, Phys. Rev. \textbf{D67} 073005 (2003)
\bibitem{sngrb} {\it e.g.} GRB GWs: M. H. van Putten {\it et al.}, Phys. Rev. \textbf{D69} 044007 (2004); SN/GRB neutrinos: S. Razzaque {\it et al.}, Phys. Rev. \textbf{D69} 023001 (2004)
\bibitem{mirabel} F. Mirabel, \texttt{arXiv:0805.2378v1} and references therein
\bibitem{ls5039} For LS5039: F. A. Aharonian {\it et al.}, J. Phys. Conf. Ser. \textbf{39} 408-415 (2006); see also S. Aiello {\it et al.}, Astropart. Phys. \textbf{28} 1–9 (2007)
\bibitem{mypaper} Th. Pradier, to be submitted to A\&A
\bibitem{gwbh} R. H. Price, Phys. Rev. \textbf{D5} 2419 (1972); see also A. Nagar {\it et al.}, Phys. Rev. \textbf{D75} 044016 (2007)
\bibitem{jetgw} E. B. Segalis \& A. Ori, Phys. Rev. \textbf{D64} 064018 (2001)
\bibitem{sgr} C. Kouveliotou {\it et al.}, Astrophys.J.\textbf{510} L115-118 (1999); see also T. Terasawa {\it et al.}, Nature \textbf{434} 1110 (2005)	
\bibitem{sgrnu} K. Ioka {\it et al.}, Astrophys. J. \textbf{633}, 1013-1017 (2005); see also F. Halzen {\it et al.} \texttt{arXiv:astro-ph/0503348v1}
\bibitem{sgrgw} J. A. de Freitas Pacheco, A\&A \textbf{396}, 397-401 (1998)
\bibitem{virgodetection2} Th. Pradier {\it et al.}, Phys. Rev. \textbf{D63} 042002 (2001)
\bibitem{virgodetection} N. Arnaud {\it et al.}, Phys. Rev. \textbf{D65} 042004 (2002)
\bibitem{aart} A. Heijboer, {\it Track reconstruction and point source searches
with Antares}, Ph.D. dissertation, Universiteit van Amsterdam (2004) \texttt{http://antares.in2p3.fr/}
\bibitem{qgexp} S. D. Biller {\it et al.}, Phys. Rev. Lett. \textbf{83}, 2108-2111 (1999); see also J. Albert {\it et al.}, \texttt{arXiv:0708.2889v1}
\bibitem{km3} U. Katz, for {\sc km3NeT}, these proceedings 
\bibitem{adv_virgo} \vo~Collab., Class. Quantum Grav. \textbf{25} 114045~(2008)


\end{thebibliography}
\end{document}